\parindent=0pt
\parskip=0.3cm
\magnification \magstephalf

\def \CM{{\hbox{\sevenrm CM}}}
\def \CdM{{\hbox{\sevenrm Cold Matter}}}
\def \max{{\hbox{\sevenrm max}}}
\def \min{{\hbox{\sevenrm min}}}
\def \gmin{{\hbox{\sevenrm gmin}}}
\def \H{{\hbox{\sevenrm Hubble}}}
\def \N{{\hbox{\sevenrm number count}}}
\def \h{\hfill}

\null
\vskip 1.5cm
\centerline{\bf DETERMINATION OF COSMOLOGY AND EVOLUTION FROM $K$-BAND}
\centerline{\bf MAGNITUDE--REDSHIFT AND NUMBER COUNT OBSERVATIONS}
\vskip 1.5cm

\centerline {by}
\vskip 0.3cm

\centerline {J.C. Jackson\footnote{*}{e-mail: john.jackson@unn.ac.uk} and Marina Dodgson}
\centerline {School of Computing and Mathematics}
\centerline {University of Northumbria at Newcastle}
\centerline {Ellison Building}
\centerline {Newcastle upon Tyne NE1 8ST, UK}
\vskip 1.2cm

\centerline{\bf ABSTRACT}
\vskip 0.3cm

We determine cosmological and evolutionary parameters from the 3CR 
$K$-band Hubble diagram and $K$-band number counts, assuming that the 
galaxies in question undergo pure luminosity evolution.  Separately
the two data sets are highly degenerate with respect to choice of
cosmological and evolutionary parameters, but in combination the
degeneracy is resolved.  Of models which are either flat or have
$\Omega_\Lambda=0$, the preferred ones are close to the canonical
case $\Omega_\CdM=1$, $\Omega_\Lambda=0$, with luminosity evolution
amounting to one magnitude brighter at $z=1$.
\vskip 0.6cm

{\bf Key words:} cosmology -- observations -- theory -- dark matter. 
\vfil\eject

{\bf 1\hskip 0.3cm INTRODUCTION}

There is now a strong concensus that the basic cosmological parameters are
known, and that we are living in a spatially flat accelerating Universe,
with $\Omega_\CM\sim 0.3$ and $\Omega_\Lambda\sim 1-\Omega_\CM\sim 0.7$ (CM
$\equiv$ Cold Matter, that is Cold Dark Matter plus the baryonic component).
This concencus is based primarily upon observations of Type Ia supernovae 
(Schmidt et al. 1998; Riess et al. 1998; Perlmutter et al. 1999), 
coupled with observations of Cosmic Microwave Background (CMB) anisotropies  
(Efstathiou et al. 1999).  Separately these observations place different 
constraints on the parameter values, but in combination degeneracies are 
removed (see also Bridle et al. 1999; Lasenby, Bridle \& Hobson 2000).  
Additional support comes from measures of the angular size of the 
acoustic horizon at decoupling, via detection of the Doppler peak 
in the CMB angular spectrum (de Bernardis et al. 2000; Hanany et al.
2000).  However, there are weaknesses in the case for support; there are
some doubts relating to the supernovae as non-evolving standard candles
(Riess et al. 1999), and the CMB angular spectrum in the range 
15' to 30' is not quite as expected (Tegmark \& Zaldarriaga 2000).
Additionally it may be that the rigid form of repulsive vacuum energy
represented by a positive cosmological constant is too restrictive, and
that more flexible forms (quintessence) might be necessary to account
for all of the observations, particularly those represented by an evolving
scalar field (Ratra \& Peebles 1988; Frieman et al. 1995; Jackson 1998a,b; 
Jackson \& Dodgson 1998; Caldwell, Dave \& Steinhardt 1998; Barrow, Bean 
\& Maguejo 2000).  Thus independent checks are always to be welcomed,
and it is in this spirit that the following considerations are offered.
We attempt to determine cosmological parameters from the $K$-band Hubble
diagram for powerful radio galaxies, treating the latter as evolving
standard candles.  The problem as posed is degenerate with respect to
choice of cosmological and evolutionary parameters; the degeneracy is
removed by coupling the Hubble diagram with $K$-band number counts.
The general idea that we cannot invoke evolution to account for the Hubble
diagram without considering the knock-on effects on the number counts, and
vice-versa, has been discussed by a number of authors (Shanks et al. 1987;
Yoshii \& Takahara 1988).  This work is an extended and updated version of
a quantitative realization of this idea given by Jackson \& Dodgson (1997;
see also Dodgson 1999).  The evolutionary model adopted is traditional
pure luminosity evolution (PLE), and in view of the bad press received by
the latter over the last few years (Zepf 1997; Kauffman \& Charlot 1998),
some justification of this assumption is required, a point to which we
shall return later.  Our aim is establish that PLE can account for the
observations in a natural way, rather than to demolish alternative
scenarios.

$K$-band work was pioneered by Lilly \& Longair (1984), who in a classic 
paper presented photometric and redshift data for an almost complete sample
of radio galaxies from the 3CR catalogue, in the range $0<z<1.6$ (see also
Lilly 1989).  The low-redshift objects in the sample are giant elliptical
galaxies, containing old stellar populations with no evidence of recent star
formation. The corresponding $K$-magnitude versus redshift relation is very
well defined, and was interpreted as indicating that the complete sample
comprises giant elliptical galaxies, created together at some formation
redshift $z_\max$, and subsequently undergoing passive stellar evolution.
Their conclusion was that these galaxies were brighter in the past, by about
one magnitude at $z=1$.  Despite a number of caveats which have arisen over
recent years, in somewhat modified form this picture has essentially 
survived.  The main controversy surrounding the use of powerful radio 
galaxies in this context centres on the question: when is a quasar not a 
quasar?  In a flux-limited sample such as 3CR, the highest redshifts 
correspond to the most powerful radio galaxies, and there is a suspicion 
that their nuclei contain low-luminosity quasars.  Lilly \& Longair were 
well aware of this possibility, and of the 77 candidates for inclusion 
in their sample, 8 were eliminated because they show quasar-like spectral 
features, particularly broad optical emission lines, and varying degrees 
of infra-red excess, associated with non-stellar emission from the active 
nuclei.  

These stringent selection criteria were designed to eliminate objects
in which nuclear activity is contributing to the $K$-band luminosity, 
either directly or indirectly.  That they fail to do so in the case of 
high-redshift sources was confirmed by discovery of the alignment effect 
(Chambers, Miley \& van Breugel 1987; McCarthy et al. 1987a; McCarthy 
et al. 1987b).  At $z>0.6$ many radio galaxies have an optical 
appearance which is very different from that of a giant elliptical, 
comprising highly elongated optical emission aligned with the radio axes, 
or a series of knots strung out along these axes (Dunlop \& Peacock 1993;
Best, Longair \& R\"ottgering 1997).  However, in the $K$-band the effect
is much less pronounced, with a contribution from an aligned rest-frame
blue component amounting to no more than about 10\% of the total
(Best, Longair \& R\"ottgering 1998: BLR); all the observational evidence
is that the underlying galaxies are giant ellipticals, and that most
of the infrared luminosity comes from an old population of passively
evolving stars (Lilly 1989; McClure \& Dunplop 2000).  BLR have presented
a revised 3CR sample with new photometry, comprising 28 galaxies with
$0.6<z<1.8$, with $K$-magnitudes corrected for any aligned and point-source
contributions.

The tightness of the $K$-band Hubble diagram has been known since
the work of Lilly \& Longair (1984) and has occasioned much debate, but
astrophysical arguments relating to this phenomenon are only just beginning
to emerge.  At low powers the radio and infrared luminosities are reasonably
well correlated, so that choosing powerful radio galaxies naturally drives
the infrared luminosity towards the top end of the galaxy luminosity
function, where there is a sharp cut-off. Thus as long as one selects
reasonably powerful radio sources, one always selects the extreme
upper end of the galaxy luminosity function (McCarthy 1993; Scarpa
\& Urry 2001).  This is in reasonable accord with the recently discovered
correlation between the mass of the elliptical bulge component of a galaxy
and that of the central supermassive black hole (Kormendy \& Richstone 1995;
Ferrarese \& Merritt 2000; Kormendy 2000), which in turn fixes the
radio power, given sufficient fuel to allow accretion at the 
Eddington limit.  However, the real puzzle is why the mass 
function of the 3CR galaxies does not evolve; BLR present evidence
that some of the most distant 3CR sources are cD galaxies in moderately
rich clusters, and a reasonable expectation is that they should continue
to accumulate mass through mergers and gas infall.  This appears to be so
in the case of brightest cluster galaxies (BCGs) (Arag\'on-Salamanca, Ellis,
Couch \& Carter 1993; Collins \& Mann 1998), to the extent that the
$K$-band BCG Hubble diagram is consistent with no luminosity evolution;
the two evolutionary effects are presumed to cancel out.  The explanation
offered by BLR is that a number of astrophysical effects conspire to
deprive the distant 3CR sources of fuel (for the central black hole)
before their masses can exceed a few times $10^{11}$ solar masses.
Low-redshift 3CR sources are different in that they are in small groups
with low velocity dispersions in which there has probably been little
mass evolution; in these sources the central black hole appears to have
been re-fuelled by a recent merger with a gas-rich galaxy.  From the
point of view of this investigation all that matters is the constancy of
the stellar mass component of 3CR galaxies over the redshift range
$0.03\leq z\leq 1.8$.

Number counts of course refer to the general population of galaxies,
although they tend to be dominated by E and S0 types (Glazebrook et al.
1994; Huang et al. 1997).  In what follows we must certainly
assume that radio-brightness is just a label, which indicates
that the corresponding galaxies are at the bright end of the $K$-band
luminosity function, but are otherwise representative members of the
general population.  There is evidence that sources need to be in the 3CR
catalogue to qualify for the $K-z$ diagram in this context; those in the
B2/6C sample examined by Eales et al. (1997) are weaker than the 3CR ones
by a factor of about 5, and for $z>0.6$ are systematically fainter in $K$,
by about 0.6 magnitudes at $z=1$.  Eales et al. (1997) argued that this
difference is due to the non-stellar component in the high-redshift objects,
but BLR show that this contribution is typically 0.1 magnitudes.

In the next section we present the $K$-band Hubble and number-count
observational data, and consider the best compromise in terms of cosmological
and evolutionary parameters, first in the case of vanishing cosmological
constant, then in the spatially flat case $\Omega_\Lambda=1-\Omega_\CM$,
and finally the general case with $\Omega_\CM$, $\Omega_\Lambda$ and $\beta$
as free parameters.  All $H_0$-dependent figures quoted here assume a
value of $100$ km sec$^{-1}$ Mpc$^{-1}$.
\vskip 0.3cm

{\bf 2\hskip 0.3cm THE $K$-BAND HUBBLE DIAGRAM AND $K$-BAND NUMBER COUNTS}

Following BLR, the $K$--$z$ sample comprises those objects classified as
Narrow Line Radio Galaxies with $0<z<0.6$ in the original 1984 sample,
except that 3C13, 3C55 and 3C263.1 were misidentified then and have been
omitted, leaving 45 objects in total.  3C13 has been transferred to the
revised high-redshift sample with a new redshift of $z=1.351$, which also
includes 6 new objects; however, 3C22 and 3C41 have been reclassified as
mini-quasars, leaving 26 radio galaxies with $0.6<z<1.8$ and a grand total
of 71 objects in the range $0<z<1.8$.  A K-correction based upon a
non-evolving spectral energy distribution for E/S0 galxies (Gardner 1992;
Gardner 1995) has been applied to each measured $K$-magnitude.  Similarly 
an aperture correction based upon a fixed physical diameter of 35 kpc has 
been applied to the measured magnitudes using the radial light profile of 
the nearby giant elliptical galaxy 3C223 (Lilly, McClean \& Longair 1984).
The apertures used are those listed in Table 1 of Lilly \& Longair (1984)
and the nominal 9 arcsec aperture corresponding to $K_{\hbox{\sevenrm corr}}$
in Table 1 of BLR.  The aperture corrections are significant only for
$z^<_\sim 0.1$, where they do not depend on cosmology, and negligible at
higher redshifts (cf. Wampler 1987). The $K$--$z$ data so corrected are
shown in Figure 1.  We assume a fixed absolute magnitude of $-24.0$ for
these galaxies, determined by the low-redshift data ($z<0.1$); evolution
will be incorporated by allowing the corresponding luminosity to evolve
as $(1+z)^\beta$, where $\beta$ is a constant to be determined by the data.
Figure 1 shows a theoretical curve with no evolution, the canonical case
$\Omega_\CM=1, \Omega_\Lambda=0, \beta=0$, with respect to which the
galaxies are clearly too bright.  It is well known that the $K$--$z$ diagram
cannot be fitted by any plausible non-evolving model (Lilly \& Longair 1984;
BLR), the best such model here being $\Omega_\CM=6.3$ (if $\Omega_\Lambda=0$).

Turning now to number counts, we have used a compilation (Metcalfe et al.
1996, and the references given there) which incorporates recent results
from the United Kingdom Infrared Telescope (UKIRT), and the Keck 10 meter 
telescope (Djorgovski et al. 1995).  The observations are shown in Figure 2.
To construct theoretical curves we have adopted a single Schechter (1976)
luminosity function $\phi(L)$

$$
\phi(L)dL=
\phi^*\left({L \over L^*}\right)^\alpha\exp\left(-{L\over L^*}\right)
d\left({L \over L^*}\right),
\eqno(1)
$$

with a cut-off luminosity $L^*$ corresponding (at $z=0$) to an absolute
magnitude $M^*=-23.6$ (Mobasher, Sharples and Ellis 1993; Glazebrook
et al. 1994; Cowie et al. 1996).  Again evolution is incorporated 
by allowing $L^*$ to evolve as $(1+z)^\beta$, where this $\beta$ and the 
one adopted for the 3CR galaxies are the one and the same.  We have allowed 
the normalization constant $\phi^*$ and the index $\alpha$ to be fixed by 
the bright and faint ends of the count data, as follows.  In the low-$z$ 
approximation the whole-sky number count $N(m)$ as a function of apparent 
magnitude $m$ follows the equation

$$
{dN \over dm}=
4\pi{\ln 10 \over 5}10^{3(m-M^*-25)/5}\phi^*\,\Gamma(5/2+\alpha).
\eqno(2)
$$

Thus the bright end ($K=12.0$, $\log dN/dm=0.33$) give gives
$\phi^*\Gamma(5/2+\alpha)=0.00665$ galaxies Mpc$^{-3}$.  The faint end
appears to be dominated galaxies with $z^<_\sim z_\max$ which are
intrinsically fainter than the cut-off luminosity, in which case the
asymptotic slope should be $-2(1+\alpha)/5$.  Matching this to the data
when $K^>_\sim 21.0$ gives $\alpha\sim -1.4$, and equation (2) then gives
$\phi^*=0.0070$ (cf. Mobasher et al. 1993).  These determinations are
reasonably well decoupled from the cosmological model.  Using one Schechter
function to describe the full galaxy population is computationally convenient,
but might appear to be an over-simplification; we shall argue that this is
not the case.  A mix of morphological types might be is used, each with its
own luminosity function; however, picking an appropriate mix is plagued 
by degeneracy problems, which is why the various $K$-band luminosity 
functions reported in the literature (Mobasher et al. 1993; Glazebrook et al. 
1994; Glazebrook et al. 1995; McCracken et al. 2000) do not agree.  
In the $K$-band the benefits of such decomposition are in any case marginal, 
because we see the old elliptical parts of galaxies, for example the nuclear 
bulge in the case of spirals (Mobasher et al. 1993; Huang et al. 1997); 
these have similar spectral energy distributions and evolutionary histories, 
and hence K-corrections which depend only weakly on morphological type.  
Thus there is little loss of generality in using a composite luminosity
function and a single K-correction; our theoretical curves incorporate the 
same E/S0 K-correction as the one used for the 3CR magnitudes.  The single 
function chosen here encodes the key features which have to be generated 
by any morphological mix, namely an intrinsically bright population 
which dominates when $K^<_\sim 18$, plus an intrinsically faint population 
with steep $\alpha$ which is dominant when $K^>_\sim 18$
(Cowie, Songaila \& Hu 1991; Gardner, Cowie \& Wainscoat 1993; Driver et al. 
1994; Gronwall \& Koo 1995; Glazebrook et al. 1995).  Despite these general 
arguments, we have checked that our results are in fact robust with 
respect to the above choice of composite function, by undertaking
representative computations using mock morphological mixes, for example 
a two-component model with $M^*=-23.6,-22.6$, $\alpha=-1.1,-1.5$, 
60\% and 40\% respectively, and a six component model based upon that 
given in Glazebrook et al. (1994), but with somewhat steeper values of 
$\alpha$, $-1.1$ for E/S0 to $-1.6$ for Im.  A formation redshift 
$z_\max=5$ has been adopted; results are not sensitive to this choice 
as long as ${z_\max}^>_\sim 4$.  Figure 2 shows the non-evolving curve 
$\Omega_\CM=1, \Omega_\Lambda=0, \beta=0$, with respect to which there 
are too many galaxies.  Searching for a good non-evolving fit with 
$\Omega_\Lambda=0$ drives the model to negative values of $\Omega_\CM$, 
as it attempts to find greater volumes.

First we concentrate on the case $\Omega_\Lambda=0$, leaving
$\Omega_\CM$ and $\beta$ as free parameters.  Two-parameter searches
certainly produce better fits to both data sets; allowing the 3CR galaxies
to be brighter in the past allows a good fit without demanding too much
of $\Omega_\CM$, and similar evolution allows more galaxies to be seen
in the critical region $15<K<20$ of Figure 2. (Beyond $K\sim 21.0$
we start to run out of bright galaxies and begin to sample the lower
reaches of the luminosity function, in part because this magnitude can
exceed the apparent magnitude corresponding to $M^*$ at $z_\max$, and
in part because most models begin to run out of volume.  There is clear
evidence from the change of slope that this is happening here.)
The problem is that the situation is highly degenerate in this respect:
there are many almost equally acceptable choices of $(\Omega_\CM,\beta)$;
the corresponding curves define standard deviations $\sigma_\H=0.47$ and
$\sigma_\N=0.13$ (in $K$ and $\log dN/dK$ respectively), which can be
used for statistical purposes.  As a rough guide to the way out of this
dilema Figure 3 shows the optimum value of $\Omega_\CM$ as a function of
$\beta$, plotted as the falling curve for the $K$--$z$ data and the rising
one for the $dN/dK$--$K$ points.  The two curves cross at $\Omega_\CM=1.1$,
$\beta=1.5$, which are thus rough estimates of the best compromise
parameters.  A firmer statistical basis is afforded by constructing a joint
$\chi^2=\chi_\H^2+\chi_\N^2$, to be minimised in a two-parameter search,
which gives $\Omega_\CM=1.11$, $\beta=1.40$.  Similarly we can construct
confidence regions, by taking the appropriate contour, in this case
$\chi^2=\chi^2_\min+5.991$ for 2 parameters and 95\% confidence.
This is shown as the continuous curve in Figure 3.  The formal one-parameter
95\% limits are $\Omega_\CM=1.11\pm 0.24$, $\beta=1.40\pm 0.11$.
In order to test for redshift dependence, the $\chi^2$ calculations have
been repeated for the smaller non-controversial 3CR sample, comprising
sources with $z<0.6$; the 95\% confidence region is shown as the dotted curve
in Figure 3.  This second region is larger than the first as expected, but
it nicely includes the former, and there is no evidence of any significant
shift with diminishing $z$.  Until relatively recently the canonical flat
CM model $\Omega_\CM=1$, $\Omega_\Lambda=0$ was much favoured, and advocates
of such a model would clearly be encouraged by these figures.  The required
luminosity evolution amounts to 1.05 magnitudes at $z=1$, which is
compatible with the predictions of population synthesis models of stellar 
evolution in elliptical galaxies (Sweigart \& Gross 1978; Lilly \& Longair 
1984; Bruzual \& Charlot 1993; BLR).

In view of the current preoccupation with more general flat models, we
have repeated the calculations with $\Omega_\Lambda=1-\Omega_\CM$
rather than zero.  The outcome is shown in Figure 4, which is not very
different from Figure 3.  There is no support here for flat accelerating
models.

Finally we approach $\Omega_\CM$, $\Omega_\Lambda$ and $\beta$ without
prejudice and treat then all as free parameters.  Inorder to display results
in the horizontal $\Omega_\CM$--$\Omega_\Lambda$ plane it is necessary to
marginalise over the third parameter, which is achieved by minimising
$\chi^2=\chi_\H^2+\chi_\N^2$ along a vertical axis corresponding to $\beta$, 
at an array of points in the horizontal plane; denoting the values so 
obtained by $\chi^2_\min(\Omega_\CM,\Omega_\Lambda)$ and the global minimum
by $\chi^2_\gmin$, the region $\chi^2_\min\leq \chi^2_\gmin+7.815$ is the
projection of the three-parameter 95\% confidence region onto the horizontal
plane; 95\% and 68\% marginalised two-parameter regions are $\chi^2_\min\leq 
\chi^2_\gmin+5.991$ and $\chi^2_\min\leq \chi^2_\gmin+2.279$ respectively 
(Press, Flannery, Teukolsky \& Vetterling 1986), which are shown in Figure 5. 
The parameters at the global minimum are $\Omega_\CM=0.31$, $\Omega_\Lambda
=-1.58$, $\beta=1.47$, corresponding to a deceleration parameter
$q_0=\Omega_\CM/2-\Omega_\Lambda=1.74$. This globally preferred model is 
indicated by the continuous curves in Figures 1 and 2.  Figure 5 shows
the lines corresponding to flatness ($\Omega_\Lambda=1-\Omega_\CM$)
and zero acceleration ($\Omega_\Lambda=\Omega_\CM/2$).  The allowed region
falls below these, where the models are open and decelerating; again there
is no support for a flat accelerating universe.  The computations have been
repeated using a sample restricted to $z<0.6$, which again show that there
is no significant $z$-dependence; for the sake of clarity we do not show
the larger confidence region.
\vskip 0.3cm

{\bf 3\hskip 0.3cm ALTERNATIVE SCENARIOS}

Although the results presented here are reasonably stable, for example
with respect to variations in the details of the luminosity function,
they would not survive much of a move away from PLE, and we must consider
the extent to which the latter is preferred, or at least tenable.
Hierarchical theories of galaxy formation, in which massive galaxies
(giant ellipticals, S0s and early type spirals) are assembled at relatively
late times from smaller components, have been much in favour of late (see
for example Kauffmann, Guiderdoni \& White 1994).  The predictions
of such theories in relation to number counts as a function of redshift
are very different from those of PLE.  Hierarchical models predict 
that the fraction of galaxies with $z>1$ should be small, whereas 
PLE predicts a tail in the distribution extending out to $z^>_\sim 2$.
Comprehensive $K$-band computations have been presented by Kauffman \&
Charlot (1998), who use this fraction as a discriminator in this
context.  Until very recently the observational evidence from redshift
surveys (Songaila et al. 1994; Cowie et al. 1996) appeared to favour 
hierarchical models.  The definitive sample is that due to Cowie et al. 
(1996), a $K\leq 20$ composite comprising two $6'\times 2'$ Hawaii Deep 
Fields (total area 26.2 arcmin$^2$) and 254 objects with spectroscopic 
redshifts.  Table 1 shows that this sample and the hierarchical predictions 
are in reasonable accord, although the $K$=16--18 result suggests that the 
latter are begining to understate the true situation.
\vskip 0.3cm

{\bf Table 1.} Fraction of galaxies with $z>1$, theoretical predictions
and observations; the former assume $\Omega_\CM=1, \Omega_\Lambda=0$,
with $\beta=0$, $\gamma=1.5$ for the hierarchical case and $\beta=1.5$,
$\gamma=0$ for PLE.
\smallbreak
\settabs 5 \columns
\+\h K mag\h &\h Hierarch\h &\h Hawaii\h &\h PLE\h &\h EES\h & \cr
\smallbreak
\+\h 16--18\h &\h  0\%\h &\h  0\%\h &\h 26\%\h &\h $>$15\%\h & \cr
\+\h 18--19\h &\h  2\%\h &\h 10\%\h &\h 48\%\h &\h $>$23\%\h & \cr
\+\h 19--20\h &\h 12\%\h &\h 12\%\h &\h 59\%\h &\h $>$28\%\h & \cr
\vskip 0.3cm

However, the latest observations tell a very different story;
Eisenhardt et al. (2000) have presented a galaxy survey to $K=20$
covering 124 arcmin$^2$, compiled by Elston, Eisenhardt \& Stanford (EES),
shown in the last column of Table 1.  Their $z>1$ criterion is based upon
colour, $J-K>1.9$ (with spectroscopic spot checks), which is believed
to be a conservative estimator, hence the lower limits.  In this sample the
$z>1$ fraction is significantly greater than the hierarchical predictions,
and at the very least compatible with PLE models.  There appears to be a
reasonable explanation for the discrepancy; Eisenhardt et al. (2000) show
that the surface density of $K$-selected $z>1$ galaxies is very clumpy,
and that the earlier samples are too small to be representative; by
mischance the latter coincide with regions in which there is a dearth
of such objects.  This is most pronounced in the case of to the
Hubble Deep Field (7 arcmin$^2$): 16 sub-fields of this extent
within the EES sample show a large cosmic variance in surface density,
from 1.0 to 6.7 galaxies/arcmin$^2$ with $J-K>1.9$ and $K<20$, with a
mean of 3.4.  The figure for the Hubble Deep Field itself is 0.6, which
is thus very deficient in such objects.  The present situation is
reminiscent of earlier $B$-band counts, when there were definitive
statements to the effect that the $z>1$ tail was missing (see for example
Kauffmann, Guiderdoni \& White 1994; Glazebrook et al. 1994), but the tail
did eventually appear (Cowie et al. 1996; see also McCracken et al. 2000).

Although we have not investigated hierarchical theories in detail,
we present one or two illustrative examples, using a simple merger
model in which $K$-luminosity per unit mass is conserved:
$\phi^*\propto (1+z)^\gamma$, $L^*\propto (1+z)^{-\gamma}$, and thus
$\phi^*L^*=$constant (see for example Koo 1990; Glazebrook et al. 1994).
All cases assume  $\Omega_\CM=1, \Omega_\Lambda=0$, but the general picture
is not sensitive to this choice.  The dashed curve in Figure 6 is the no
evolution case $\gamma=0$.  A long-standing problem is the excess counts
which become apparent as $K$ exceeds 14, too close for cosmological
geometry to have any effect.  The merger model is counter-productive
in this respect; although there are more galaxies at fainter magnitudes,
they are intrinsically fainter and hence nearer, and the reduced volume
cancels any benefit.  Demergers are effective only when $K^>_{\sim} 20$,
where $L<L^*$ galaxies are dominant.  The continuous curve in Figure 6
is the case $\gamma=1.5$; the fit is reasonable good at both ends but poor
at intermediate magnitudes.  The `Hierarch' column in Table 1 corresponds
to this model; the figures are remarkably close to those produced by
Kauffmann \& Charlot (1998).  The fit looks better with normalization at some
intermediate magnitude (McCracken et al. 2000), but there is then a deficit
of galaxies at the bright end for which there is no reasonable explanation.
A local hole in the galaxy distribution has been suggested (Shanks 1989;
Huang et al. 1997), but there appear to be insuperable problems with this
idea (McCracken et al. 2000).

As a further possibility we have considered mergers plus luminosity
evolution, $\phi^*\propto (1+z)^\gamma$, $L^*\propto (1+z)^{\beta-\gamma}$;
with $\beta=\gamma$ the luminosity per galaxy is now constant. The case
$\beta=\gamma=1.5$ is shown by the dash-dotted curve in Figure 6,
which goes some way towards generating the excess intermediate counts,
but over-predicts those at the faint end. These effects are well-known,
and have been discussed in the literature in various guises (Broadhurst,
Ellis \& Glazebrook 1992; Glazebrook et al. 1994; McCracken et al. 2000;
see also Driver et al. 1998 for related I-band considerations).
If $\beta$ and $\Omega_\CM$ are given free rein, with $\gamma=1.5$ and
$\Omega_\Lambda=0$, high values of $\beta$ and $\Omega_\CM$ are chosen,
the latter to reduce the volume available at high $z$.  Of those
combinations allowed by the Hubble diagram (the dashed curve in Figure 3),
the choice $\beta=1.3$ and $\Omega_\CM=1.8$ gives a good fit, but would
hardly be regarded as acceptable.

Apart from considerations relating to $N(>z)$, alternatives to PLE 
create more problems than they solve.  If the EES picture is accepted 
and confirmed by similar surveys, then the balance of favour must move 
away from hierarchical models. At the very least we can say that the 
obituries published over recent years with regard to PLE models were 
premature.
\vskip 0.3cm

{\bf 4 \hskip 0.3cm CONCLUSIONS}

The main conclusion is surprisingly conservative: the Lilly \& Longair/
BLR scenario is reaffirmed, and of the models which have $\Omega_\Lambda=0$
or are flat, the best one is close to the canonical CM case, with
$\Omega_\CM=1$ and luminosity evolution represtented by $(1+z)^\beta$ with
$\beta\sim 1.4$.  This is the model favoured by BLR on the basis of the 3CR
observations alone.  The significant point to be made here is that the
same model fits the number count observations; pure luminosity evolution
accounts for the apparent excess of $K$-band galaxies, and density evolution
need not be invoked.  It is instructive to look for qualitative aspects of 
the data which fix this scene.  There is very little to choose between the 
models represented by the Hubble $\Omega_\CM(\beta)$ curve of Figure 3,  
but the degeneracy is removed by fixing $\beta$.  With respect to number 
counts the situation is less degenerate; the excess of galaxies becomes 
significant as $K$ exceeds 15, which is too close to be accounted for by 
cosmological geometry; it is this aspect of the data which insists upon
luminosity evolution and fixes $\beta\sim 1.4 \hbox{ to }1.5$.  The
statistical considerations are just a refinement of this robust general
picture.

If no restriction is placed upon $\Omega_\Lambda$ a somewhat different
picture emerges; the preferred models are open and decelerating, although
the canonical CM case is just about allowed.  In qualitative terms this
is a robust conclusion, and it is easy to see why it is favoured by the
data.  With no evolution the $K$-band Hubble diagram favours closed models
and the large decelerations normally associated with high densities,
for example $q_0=3.1$ if $\Omega_\Lambda=0$.  Conversely the large numbers
of galaxies revealed by $K$-band number counts favour open models and
the small decelerations normally associated with low densities.
In conjunction with the moderating effects of luminosity evolution,
these conflicting requirements are reconciled in the open decelerating
confidence region indicated in Figure 4.

There is no support here for the low-density flat accelerating models
currently much in favour.  Indeed it would be difficult to reconcile
these with the 3CR observations alone; the problem is illustrated by
the non-evolving $\Omega_\CM=0.3$, $\Omega_\Lambda=0.7$ curve in Figure 1.
The luminosity evolution required to reconcile this curve with the data
amounts to 1.7 magnitudes at $z=1$, which would greatly over-predict the
number counts, and would in any case almost certainly be ruled out by
models of stellar evolution.  Similar considerations allowed BLR to
eliminate low-density $\Omega_\Lambda=0$ models on the basis of the
3CR observations alone.  We refrain from speculation about the discrepancy
between the results presented here and those arising from the Type Ia
supernova/CMB observations, except to refer the reader to the remarks made
in the introduction.  Significant differences are that our results derive
entirely from relatively nearby observations (i.e. $z^<_\sim 2$, rather than
the decoupling redshift $z\sim 1000$), and that we attempt to determine
evolution, rather than assuming its absence.
\vskip 0.6cm

{\bf ACKNOWLEDGMENTS}

Marina Dodgson acknowledges receipt of an University of Northumbria 
internal research studentship.  It is a pleasure to thank Tom Shanks
and Nigel Metcalfe of the University of Durham and John Gardner of the
NASA Goddard Space Flight Centre for their advice and constructive
criticism.
\vskip 0.6cm
\vfil\eject

{\bf FIGURE CAPTIONS}

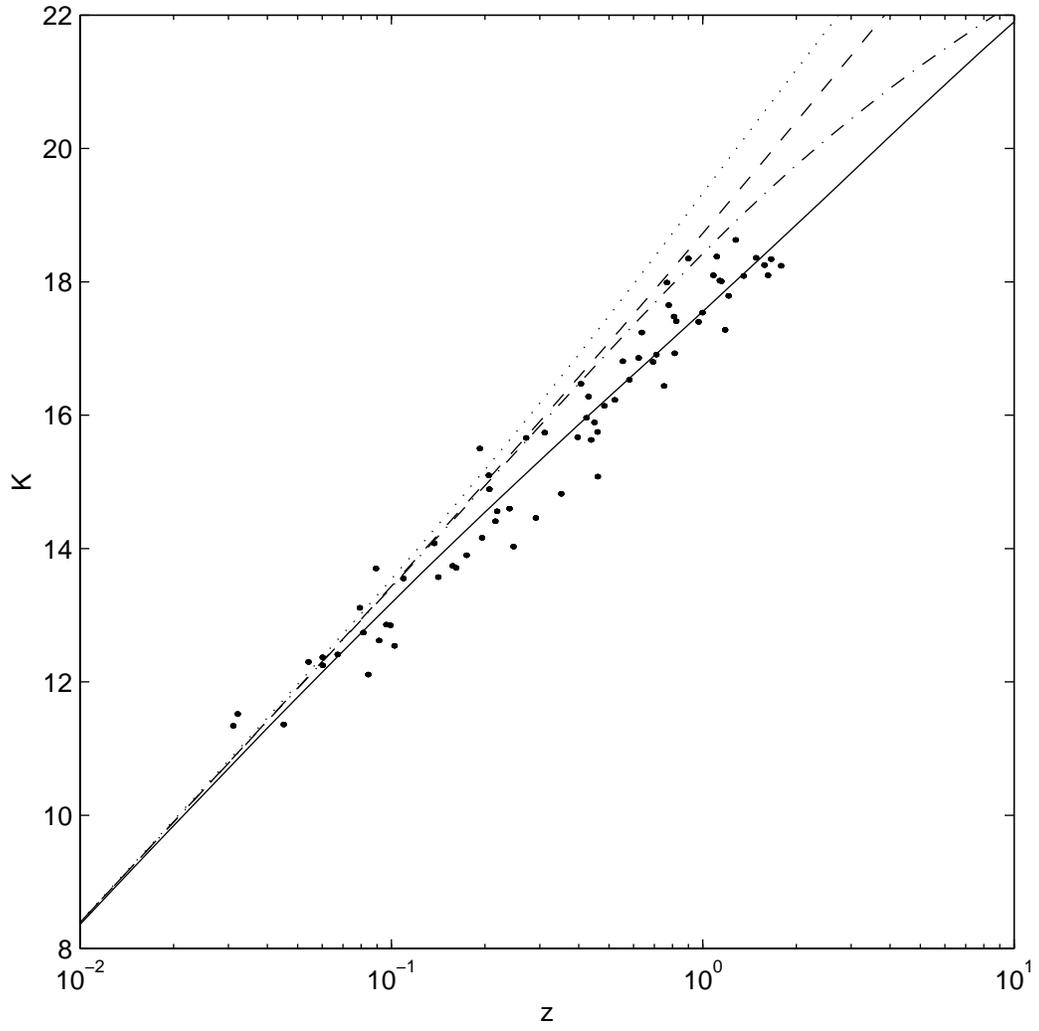
Figure 1.  The $K$-band Hubble diagram for 3CR radio galaxies according
to Best, Longair \& R\"ottgering (1998).  The dashed curve is the 
canonical case $\Omega_\CM=1$, $\Omega_\Lambda=0$, with no luminosity 
evolution.  The dotted curve is the flat accelerating case with
$\Omega_\CM=0.3$, $\Omega_\Lambda=0.7$, again non-evolving.  The other
curves allow luminosity evolution, and are determined by the Hubble and
number count data in combination.  The continuous curve is the global
optimum $\Omega_\CM=0.3$, $\Omega_\Lambda=-1.6$, $\beta=1.5$.  The
dash-dotted curve is the best compromise with the cosmological parameters
fixed at the flat accelerating values $\Omega_\CM=0.3$, $\Omega_\Lambda=0.7$,
for which the optinum value of $\beta$ is 1.2.

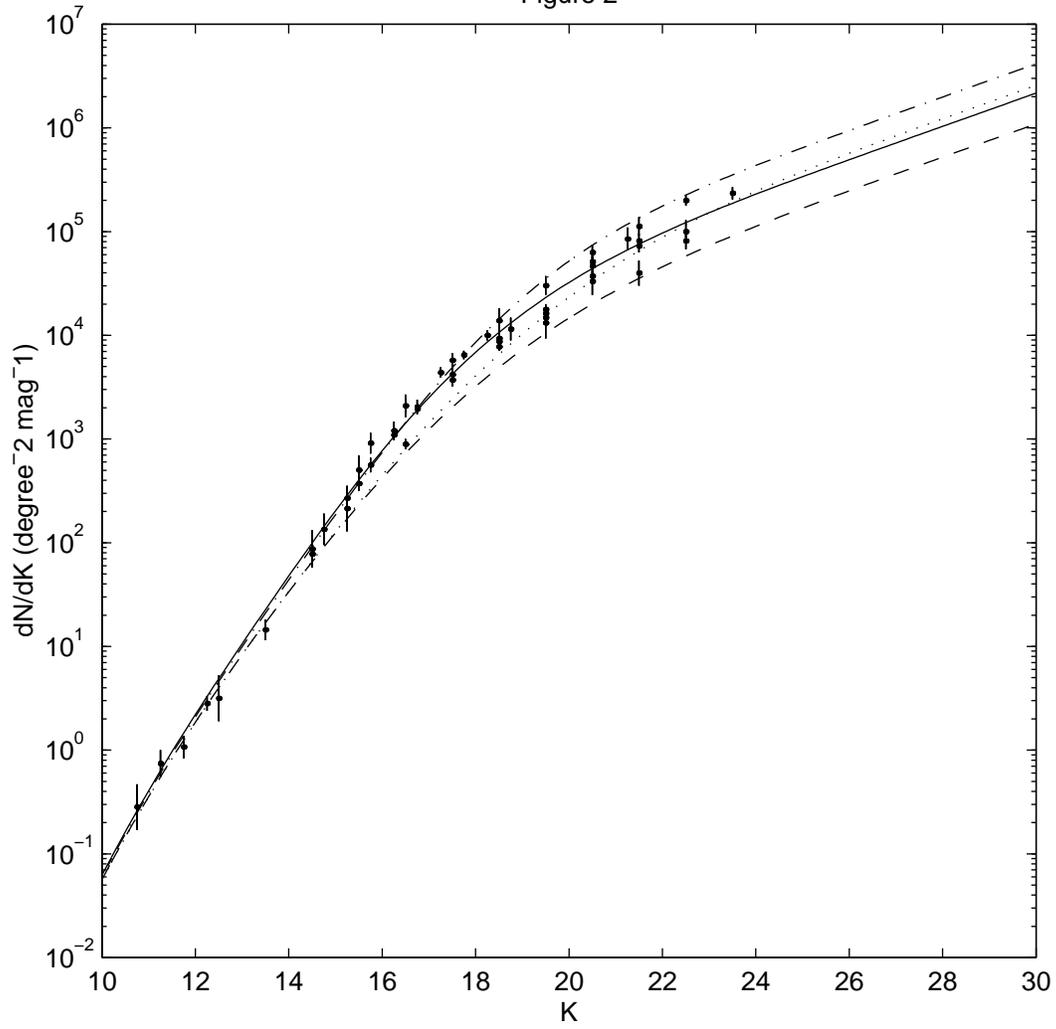
Figure 2.  $K$-band number counts according to Metcalfe et al. (1996).
The various curves are as in Figure 1.

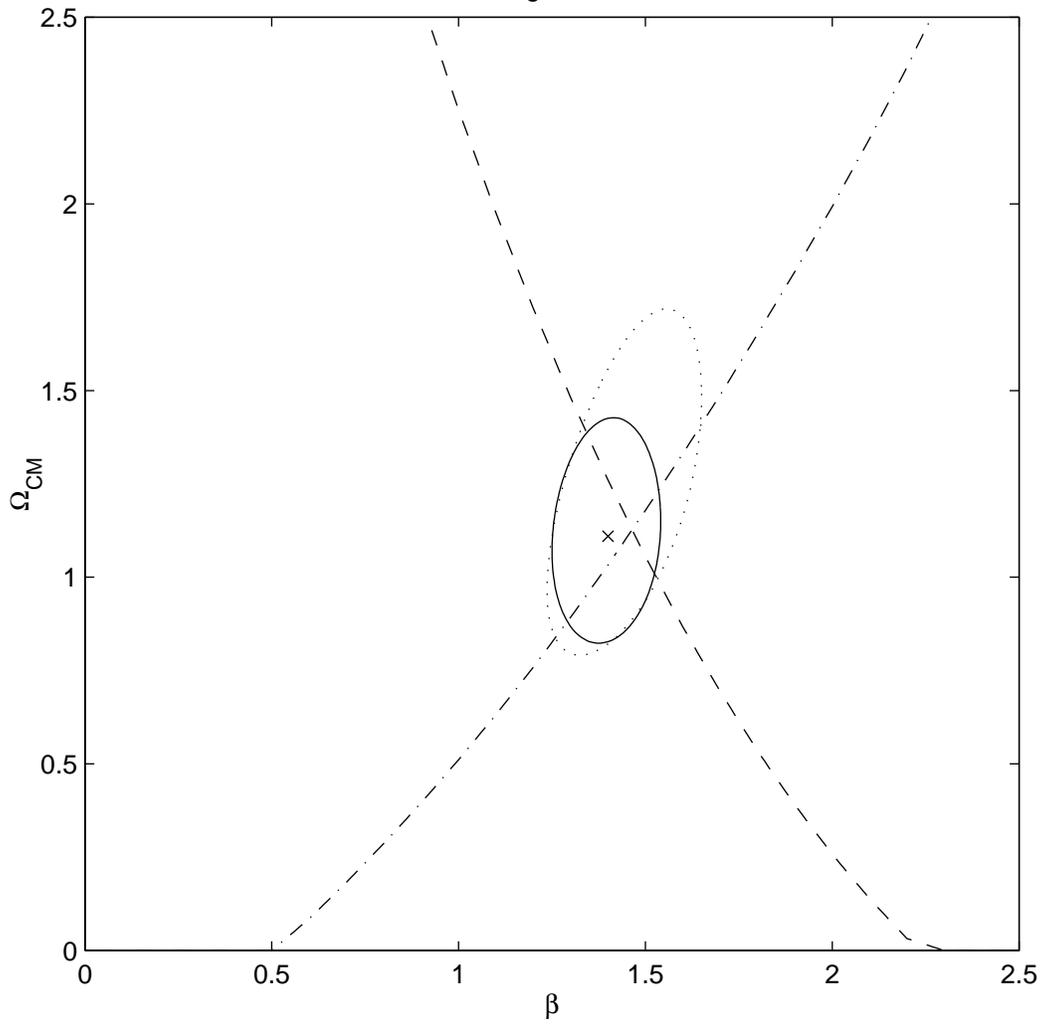
Figure 3.  Best-fit values of $\Omega_\CM$ as a function of evolutionary
parameter $\beta$, assuming $\Omega_\Lambda=0$.  The dashed and dash-dotted
curves are determined separately by the Hubble and number count data 
respectively.  The two curves cross at $\Omega_\CM=1.1$, $\beta=1.5$. 
The cross indicates the optimum parameters $\Omega_\CM=1.11$, $\beta=1.40$,
determined by the Hubble and number count data in combination; the
continuous curve is the corresponding 95\% confidence region.  These
results refer to the full 3CR sample with $0.03<z<1.8$; the dotted contour
corresponds to a low-redshift sub-sample with $z<0.6$.

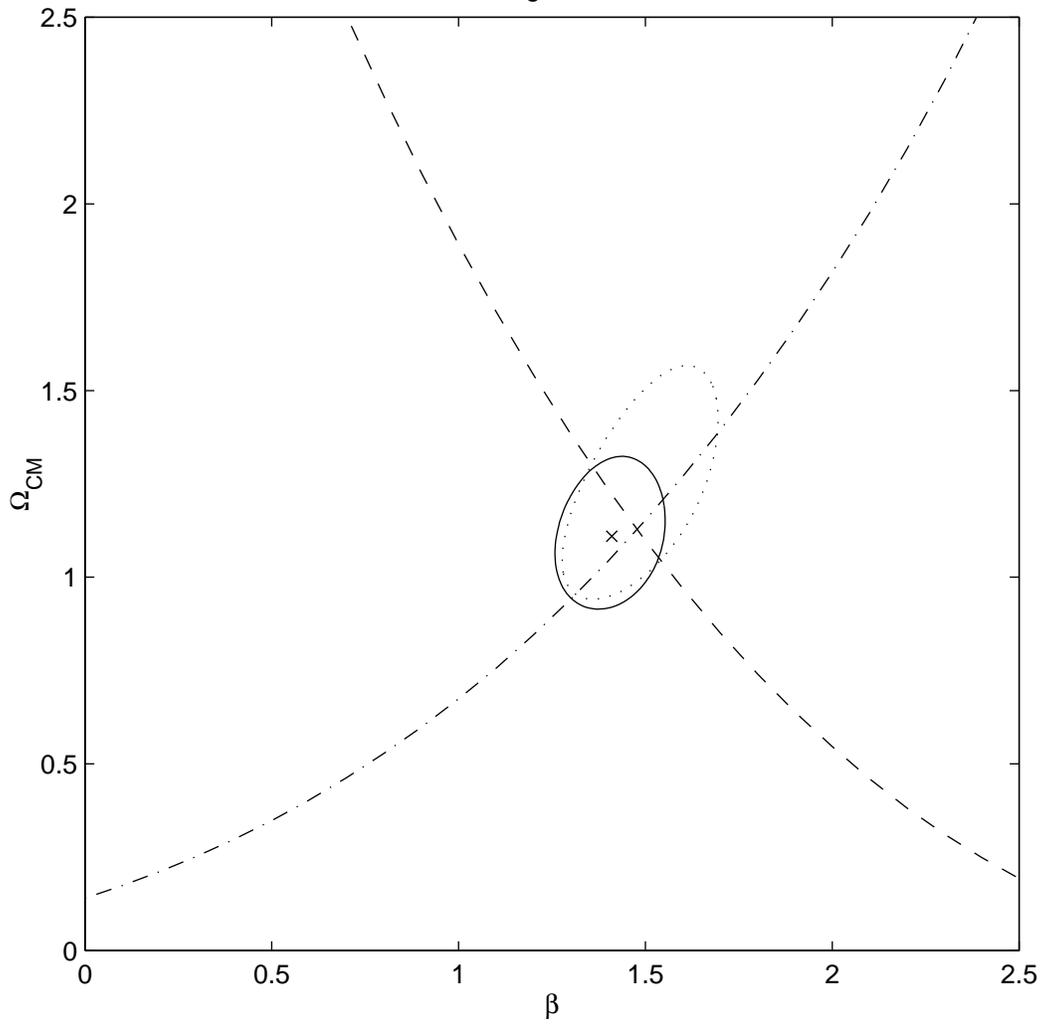
Figure 4.  Best-fit values of $\Omega_\CM$ as a function of evolutionary
parameter $\beta$, assuming $\Omega_\Lambda=1-\Omega_\CM$.  The various
curves are as in Figure 3.  The cross indicates the optimum parameters
$\Omega_\CM=1.11$, $\beta=1.41$, determined by the Hubble and number
count data in combination.

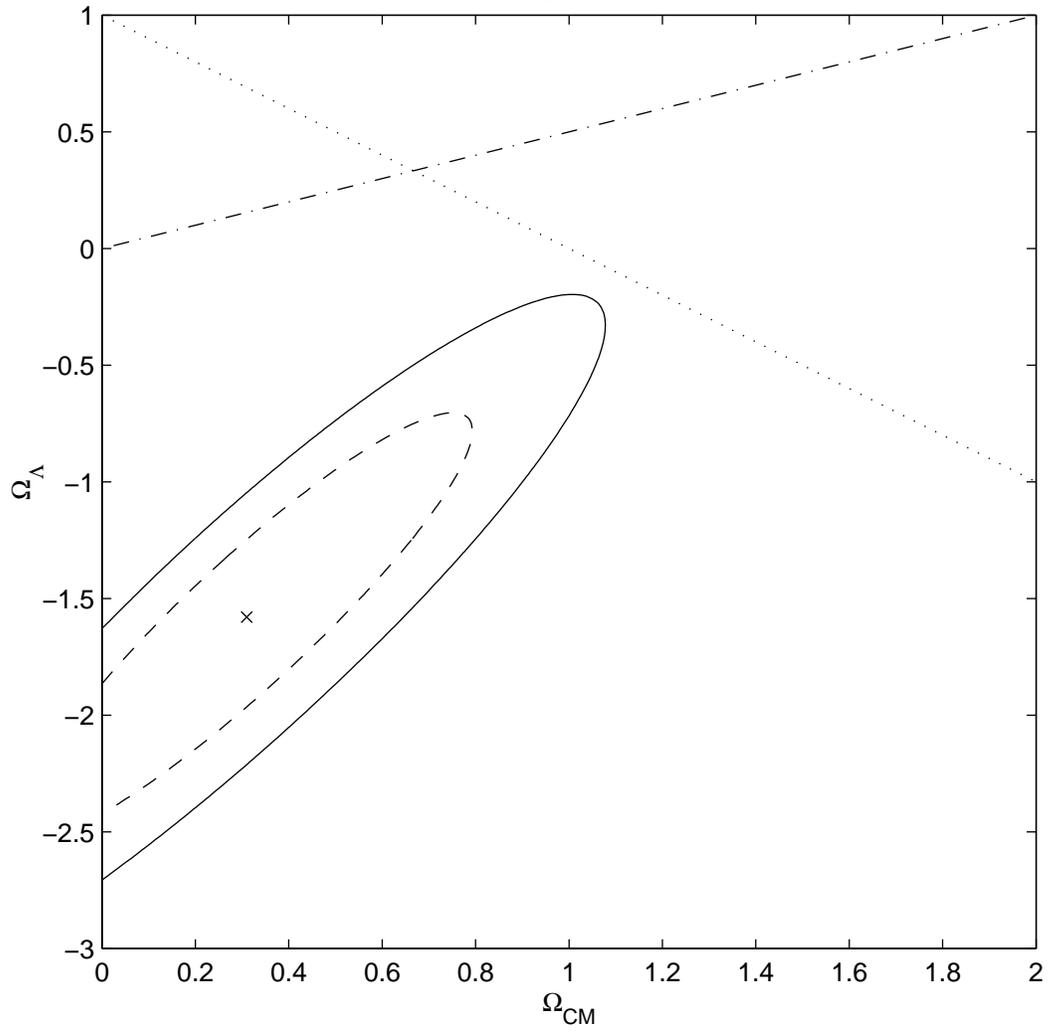
Figure 5.  Confidence regions (95\% and 68\%) in the
$\Omega_\CM$--$\Omega_\Lambda$ plane determined by the Hubble and
number count data in combination, marginalised over $\beta$.  The cross
indicates the optimum parameters $\Omega_\CM=0.31$, $\Omega_\Lambda=-1.58$,
$\beta=1.47$.  The dash-dotted line corresponds to zero acceleration,
and the dotted one to flatness.

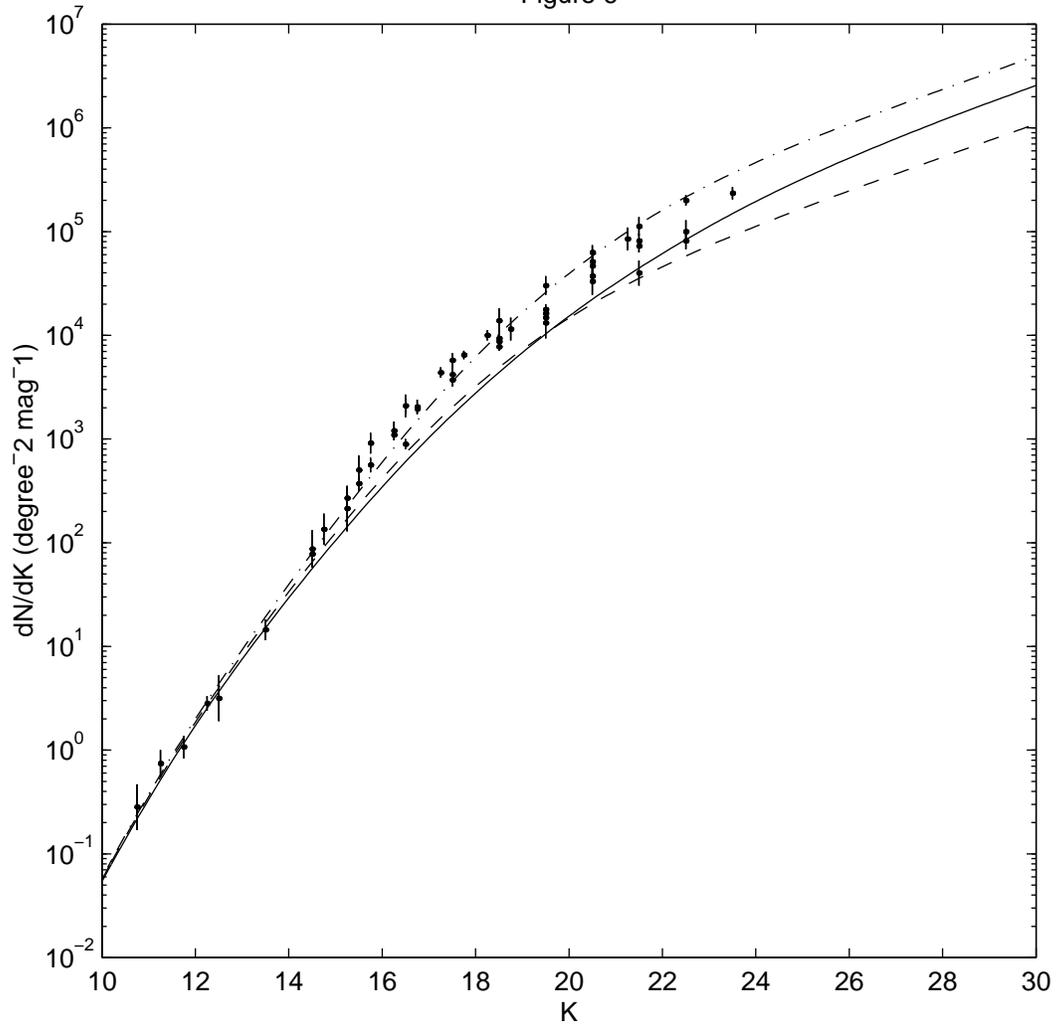
Figure 6.  $K$-band number counts according to alternatives to PLE.
The continous curve corresponds to a simple merger model, and the dash-dotted 
curve to a mix of mergers and luminosity evolution; for comparison the 
dashed curve shows the no evolution case.  See text for further details.
\vfil\eject

{\bf REFERENCES}

Arag\'on-Salamanca A., Ellis R., Couch W.J., Carter D., 1993,
MNRAS, 262, 764

Barrow J., Bean R., Maguejo J., 2000, MNRAS, 316, L41

Best P.N., Longair M.S., R\"ottgering H.J.A., 1997, MNRAS, 292, 758

Best P.N., Longair M.S., R\"ottgering H.J.A., 1998, MNRAS, 295, 549 (BLR)

Bridle S.L., Eke, V.R., Lahav O., Lasenby A.N., Hobson M.P., Cole S.,
Frenk C.S., Henry J.P., 1999, MNRAS, 310, 565

Broadhurst T.J., Ellis R.S., Glazebrook K., 1992, Nat, 355, 55

Bruzual A.G., Charlot S., 1993, ApJ, 405, 538

Caldwell R.R., Dave R., Steinhardt P.J., 1998, Phys. Rev. Lett., 80, 1582

Chambers K.C., Miley G.K., van Breugel W.J.M., 1987, Nature, 329, 624

Cowie L.L., Songaila A., Hu E.M., 1991, Nat, 354, 460

Cowie L.L., Songaila A., Hu E.M., Cohen J.G., 1996, AJ, 112, 839

Collins C.A., Mann R.G., 1998, MNRAS, 297, 128

de Bernardis P. et al., 2000, Nature, 404, 955

Djorgovski S. et al., 1995, ApJ, 438, L13

Dodgson M., 1999, Ph.D. thesis, University of Northumbria

Driver S.P., Phillipps S., Davies J.I., Morgan I., Disney M.J., 1994,
MNRAS, 268, 404

Driver S.P., Fern\'andez-Soto A., Couch W.J., Odewahn S.C., 
Windhorst R.A., Phillipps S., Lanzetta K., Yahil A., 1998, ApJ, 496, L93

Dunlop J.S., Peacock J.A., 1993, MNRAS, 263, 936

Eales S., Rawlings S., Law-Green D., Cotter G., Lacy M., 1997,
MNRAS, 291, 593

Efstathiou G., Bridle S.L., Lasenby A.N., Hobson M.P., Ellis R.S., 1999,
MNRAS, 303, 47

Eisenhardt H.J., Elston R., Stanford S.A., Dickinson M., Spinrad H.,
Stern D., Dey A., 2000, e-print, astro-ph/0002468

Ferrarese L., Merritt D., ApJ, 539, L9

Frieman J.A., Hill C.T., Stebbins A., Waga I., 1995,
Phys. Rev. Lett., 75, 2077
                                     
Gardner J.P., 1992, Ph.D. thesis, University of Hawaii

Gardner J.P., 1995, ApJS, 98, 441

Gardner J.P., Cowie L.L., Wainscoat R.J., 1993, ApJ, 415, L9

Glazebrook K., Peacock J.A., Collins C.A., Miller L., 1994, MNRAS, 266, 65

Glazebrook K., Peacock J.A., Miller L., Collins C.A., 1995, MNRAS, 275, 169

Gronwall C., Koo D.C., 1995, ApJ, 440, L1

Hanany S. et al., 2000, ApJ Letters, 545, 5

Huang J.S., Cowie L.L., Gardner J.P., Hu E.M., Songalia A., Wainscoat R.J.,
1997, ApJ, 476, 12

Jackson J.C., 1998a, MNRAS, 296, 619

Jackson J.C., 1998b, Mod. Phys. Lett. A, 13, 1737

Jackson J.C., Dodgson M., 1997, in The Hubble Space Telescope and 
the High-Redshift Universe,\hfil\break
37th Herstmonceux Conference, World Scientific, Singapore, p. 149 

Jackson J.C., Dodgson M., 1998, MNRAS, 297, 923

Kauffmann G., Guiderdoni B., White S.D., 1994, MNRAS, 267, 981

Kauffmann G., Charlot S., 1998, MNRAS, 297, L23

Kormendy J., Richstone D., 1995, ARA\&A, 33, 581

Kormendy J., 2000, Science, 289, 1484

Lasenby A.N., Bridle S.L., Hobson, M.P., 2000, 
Astrophys. Lett. \& Communications, 37, 327

Lilly S.J., 1989, ApJ, 340, 1989

Lilly S.J., Longair M.S., 1984, MNRAS, 211, 833

Lilly S.J., McClean I.S., Longair M.S., 1984, MNRAS, 209, 401

McCarthy P.J., Spinrad H., Djorgovski S., Strauss M.A., 
van Breugel W.J.M., Liebert J., 1987a, ApJ, 319, L39

McCarthy P.J., van Breugel W.J.M., Spinrad H., Djorgovski S., 
1987b, ApJ, 321, L29

McCarthy, P.J., 1993, Ann. Rev. Astron. Astrophys, 31, 639

McClure R.J., Dunlop J.S., MNRAS, 317, 249, 2000

McCracken H.J., Metcalfe N., Shanks T., Campos A., Gardner J.P.,
2000, MNRAS, 311, 707

Metcalfe N., Shanks T., Campos A., Fong R., Gardner J.P., 1996, 
Nature, 383, 236

Mobasher B., Sharples R.M., Ellis R.S., 1993, MNRAS, 263, 560

Ratra B., Peebles P.J.E., 1988, Phys. Rev., D37, 3406

Perlmutter S. et al., 1999, ApJ, 517, 565

Press W.H., Flannery B.P., Teukolsky S.A., Vetterling W.T., 1986,
Numerical Recipes, Cambridge University Press, Cambridge, pp. 532-536 

Riess A.G. et al., 1998, AJ, 116, 1009

Riess A.G., Filppenko A.V., Weidong L., Schmidt B.P., 1999, AJ, 118, 2668

Scarpa R., Urry C.M., 2001, ApJ, in the press (astro-ph/0104183) 

Schmidt B.P. et al., 1998, ApJ, 507, 46

Shanks T., Couch W.J., McHardy I.M., Cooke B.A., Pence W.D., 1987, 
in Bergeron et al., eds, High Redshift and Primeval Galaxies. 
Editions Fronti\'ers, Paris, p. 197

Shanks T., 1989, in Bowyer S., Leinert C., eds, Proc. IAU Symp. 139,
Galactic and Extragalactic Background Radiation.  Kluwer, Dordrecht, p. 269

Songaila A., Cowie L.L., Hu E.M., Gardner J.P., 1994, ApJS, 94, 461

Sweigart A.V., Gross P.G., 1978, ApJ Suppl., 36, 405

Tegmark M., Zaldarriaga M., 2000, Phys. Rev. Lett., 85, 2240

Wampler E.J., 1987, A \& A, 178, 1

Yoshii Y., Takahara F., 1988, ApJ, 326, 1

Zepf S.E., 1997, Nat 390, 377

\bye